# Trapping and hopping of bipolarons in DNA: Su-Schrieffer-Heeger model calculations

J. H. Wei,[1,*] L. X. Wang,[1] K. S. Chan,[2] and YiJing Yan[3]

[1] *Department of Physics, Shandong University, Jinan, China*
[2] *Department of Physics and Material Science, City University of Hong Kong, Hong Kong*
[3] *Department of Chemistry, Hong Kong University of Science and Technology, Kowloon, Hong Kong*

With the Su-Schrieffer-Heeger model involving the effects of solvent polarization and external electric field, we show that bipolaron maybe more stable than two polarons when a dication induced into a DNA stack. Under the high electric field, the dication can move quite a long distance through the DNA by a series of hopping process, partially losing its configuration instantaneously due to the nonadiabatic effects.

Long-range charge transport in duplex DNA is of great importance for its relevance to both the blueprint of life [1, 2] and the potential applications in molecular electronics [3, 4]. It is found experimentally that a radical cation (hole) introduced into a duplex DNA oligomer in solution can migrate over distance before being trapped by reactions with water and/or molecular oxygen [5–7]. To explain the long-range charge transport in DNA, a phonon-assisted polaron-like hopping mechanism was put forward [8]. In that view, the radical cation is stabilized with a distortion of the DNA and its environment (water molecules and counterions) [9]. This model is supported by the detailed electrical transport measurements through DNA molecules [10], in which the conduction is due to the thermal motion of small polarons. *Ab initio* calculations have also provided clear evidences for the formation of small polaron involving the change of chemical bonds in each base pair as well as the positions of hydrogens and oxygens [11]. With a Su-Schrieffer-Heeger (SSH) model similar to that used for conjugated polymers [12, 13], Conwell and Rakhmanova have showed that large polarons extended over 5-7 base pairs may exist in DNA [14]. They have also found that the time required for polaron formation is of the order of picosecond [15]. Bishop *et al.* recently studied the normal modes of polaron solutions to the Peyrard-Bishop-Holstein (PBH) model for DNA polymers [16].

What will happen when there are more than one charge, such as two cations or a dication, introduced into a duplex DNA oligomer? It has been argued that, as a result of the strong electron-lattice interaction, two polarons on a $\pi$-conjugated chain coalesce into a bipolaron, which is a pair of like charges with an associated strong local geometry deformation [17]. Whether a DNA double strand structure supports a bipolaron or two polarons needs theoretical clarification because it is an essential question directly relevant to experiments and applications.

In this paper, we investigate the dication (2 holes) trapping and hopping properties in a DNA molecule. We shall exploit the celebrated SSH model to incorporate the double-chain interaction in order to describe the DNA double strands of Watson-Crick (W-C) base pairs, G/C, C/G, A/T, and T/A. The SSH model has shown a remarkable success in the study of the electronic conductivity and optical phenomena in $\pi$ systems. To include the effects of solvent polarization and external electric field on the DNA charge transfer, we modify the standard SSH model with two additional terms; i.e.,

$$H = H_S + H_E + H_F. \tag{1}$$

The first term $H_S$ is the extended, double-strand SSH-like Hamiltonian:

$$\begin{aligned} H_S \;=\; & \sum_{j,n}\{\epsilon_{j,n}c^{+}_{j,n}c_{j,n} + \frac{M}{2}\dot{u}^2_{j,n} + \frac{K}{2}(u_{j,n+1}-u_{j,n})^2 \\ & -[t_0 - \alpha(u_{j,n+1}-u_{j,n})](c^{+}_{j,n}c_{j,n+1} + \text{H.c.})\} \\ & + \sum_n [t_s c^{+}_{\text{I},n}c_{\text{II},n} + \text{H.c.}]. \end{aligned} \tag{2}$$

Here, $j$ = I, II denoting the strand index, $c^{+}_{j,n}$ ($c_{j,n}$) is the creation (annihilation) operator of an electron at the $n^{\text{th}}$ base of strand $j$ with the on-site energy of $\epsilon_{j,n}$, while $u_{j,n}$ stands for the displacement of the base from its equilibrium position. The on-site energy parameters are chosen from Ref. 18: $\epsilon_{\text{T}}$ is taken as the zero of our energy scale, then $\epsilon_{\text{C}} = 0.21$ eV; $\epsilon_{\text{A}} = 0.9$ eV and $\epsilon_{\text{G}} = 1.39$ eV. The other intrastrand interaction and phonon parameters are chosen from Ref. 14; i.e., the zero-displacement hopping integral $t_0 = 0.3$ eV, the electron-phonon coupling $\alpha = 0.6$ eV/Å, the phonon spring constant $K = 0.85$ eV/Å$^2$ as inferred from the measured sound velocity along the stacks, and $M = 4.35\times10^{-22}$g taken to be the base pair mass. The interstrand interaction described by the last term of Eq. (2) is assumed to be dominant by the W-C base pair coupling. According to the *ab initio* calculations [19, 20], the base pair coupling strength is on the order of the intrastrand coupling, so we choose $t_s = t_0$ in the following numerical study.

The second term in Eq. (1) describes the mean field electron-electron ($e$-$e$) interaction, which is screened in the solvent by the polarizable and/or counterion surroundings that are crucial for the stability of DNA molecules [3, 21, 22]. Even under the dry condition there are still some counterions from the synthesizing procedure. So we adopt the screened Coulomb potential (SCP) continuum solvent model [23],

$$H_E = \frac{1}{4\pi}\sum_{j,n}\Big\{\frac{\Delta w_{j,n}}{R_{j,n}}\Big[\frac{1}{D(R_{j,n})}-1\Big] + \sum_{(j',m)\neq(j,n)}\frac{\Delta w_{j',m}}{r_{j,n;j',m}D(r_{j,n;j',m})}\Big\}c^{+}_{j,n}c_{j,n} \quad (3)$$

where $R_{j,n}$ is the effective size of base pair $n$ at strand $j$. $r_{j,n;j',m}$ is the distance between base$(j,n)$ and $(j',m)$, and only interaction between the nearest neighbor base pairs is concerned in the present work. $\Delta w_{j,n} \equiv \langle c^{+}_{j,n}c_{j,n}\rangle_0 - \langle c^{+}_{j,n}c_{j,n}\rangle$ gives the change of the electron mean occupation number on the specified strand-base site upon the hole injection into the initially charge-neutral DNA molecule. In the present work, $\langle c^{+}_{j,n}c_{j,n}\rangle_0 = 2$ to allow each site of the possibility with a dication or two holes, by assuming there are two active, spin-free, degenerate states involved in each $(j,n)$-site. $D(r)$ is the screening function and related to the dielectric function $\epsilon(r)$ by

$$\epsilon(r) = D(r)\Big[1+\frac{r}{D(r)}\frac{d}{dr}D(r)\Big]^{-1}. \quad (4)$$

*Electronic address: wjh@sdu.edu.cn

For the DNA sequence-based structure, $\epsilon(r)$ can be determined by the analytical form [24],

$$\epsilon(r) = 78.3 - 77.3(r/2\kappa)^{\eta}/(\sinh(r/2\kappa))^{\eta} \quad (5)$$

where $\kappa$ and $\eta$ are adjusting variables. $\epsilon(r)$ yields dielectric profiles with sigmoidal behavior in $r$ and so does $D(r)$.

We can recast Eq. (3) to a familiar mean-field $e$-$e$ interaction form,

$$H_E = \sum_{j,n}(U_c\Delta w_{j,n} + \sum_{(j',m)\neq(j,n)} V_c\Delta w_{j',m})c^{+}_{j,n}c_{j,n}. \quad (6)$$

Obviously, if $\epsilon(r)=\epsilon_0$ one gets $D(r)=\epsilon_0$, then Eq. (6) returns to the non-screened $e$-$e$ interaction, in which case a reasonable estimation in DNA molecular gives the on-site $e$-$e$ interaction $U_0 \sim t_0$ and the nearest neighbor one $V_0 \sim 0.2t_0$. When screening effects concerned, after solving Eq. (4) and Eq. (5) with reasonable parameters we find for DNA molecular in water solution $U_c \sim -1.5U_0$ and $V_c \sim 0.04V_0$. It means that the Hubbard repulsion between two like charges may be turned into an attractive interaction by solvent polarization in the large screened limit.

The last term $H_F$ in Eq. (1) describes the interaction of DNA with the external electric field, treated in the Coulomb gauge with a scalar potential and the dipole approximation. It is to adopt [25, 26]

$$H_F = e\mathcal{E}\sum_{j,n}[(n-1)a + u_{j,n}](c^{+}_{j,n}c_{j,n} - \langle c^{+}_{j,n}c_{j,n}\rangle_0). \quad (7)$$

Note that the characteristic field strength for the given DNA parameters is $\mathcal{E}_0 \equiv 2\hbar\omega_0/(ea) = 5.8\times10^5$ V/cm, where the prefactor 2 is taken into consideration of double strands, $\omega_0 = (K/M)^{1/2}$ is the phonon frequency, and $a = 3.4$ Å is the distance between two adjacent base pairs [15].

We start with the stationary solution to the system [Eqs. (1)–(7)] after two hole charges introduced into an initially charge-neutral DNA molecule in the equilibrium configuration. A self-consistent method similar to that in Ref. 27 is used to determine the stabilized lattice configuration by minimizing the total energy under the fixed-ends condition. The hole-charge and lattice-distoration distributions will be reported, respectively, in terms of

$$Q^h_n \equiv -\sum_j(\langle c^{+}_{j,n}c_{j,n}\rangle - \langle c^{+}_{j,n}c_{j,n}\rangle_0), \quad (8a)$$

$$y_n \equiv \sum_j(u_{j,n+1}-u_{j,n}). \quad (8b)$$

For demonstration, we first randomly select a DNA sequence with a total length of $N = 28$ base pairs, containing some singlet and multiplet G/C (or C/G) pair(s), separated by A/T- or T/A-bridges. Fig. 1 shows the electron spectrum corresponding to the equilibrium lattice configuration in the absence of external field and $e$-$e$ interaction, where (a), (b), and (c) illustrate the changes of electron spectrum with the doping by 0, 1, and 2 holes, respectively. Consistent with the findings of Conwell *et. al.*[14], the half-occupied polaron level is heightened by about 0.35 eV for the random arrangement of base pairs. While the unoccupied bipolaron energy level is pulled further into the band gap, by about 0.72 eV higher than the neutral system, as shown in Fig. 1c.

The resulting stationary hole and lattice distributions over individual base pairs are reported in Fig. 2a and Fig. 2b, separately, studied within three cases: without $e$-$e$ interaction$(U,V=0)$, with non-screened$(U,V=U_0,V_0)$ and with screened $e$-$e$ interaction$(U,V=U_c,V_c)$. All of these figures show that the two holes introduced into the DNA do not separate each other to form two polarons; they rather bind together and be accompanied by the lattice distortion to form a stable bipolaron localized in the G/C-quadriplex for its lowest ionization potential (IP) among all base pairs in the given sequence. In fact, the bipolaron remains stable also at a G/C-singlet site; see Fig. 2a's insert for a sequence contains no G/C-multiplet. (It would imply that in our model the effective second IP of G/C is still lower than the first IP of A/T.) The extend of the bipolaron wave function and the lattice distortion is found to be about 3 to 5 base pairs, a little narrower than that reported for a single polaron (about 5-7 base pairs) [14]. In all cases we studied, the bipolaron prefers to self-trap around the minimum free-energy site of the total DNA system if there is no external electric field applied.

To account for the above observed stability of bipolaron rather than two polarons, let us recalled some es-tablished results in a quasi-one dimensional system described with SSH model [17]. Upon the formation of an excitation (soliton, polaron or bipolaron) by introduce excess charges, the total free-energy change consists of mainly (i) the increase of the distortion energy, (ii) the decrease of ionization energy, and (iii) the increase of the Coulomb repulsion between similar charges. The ionization energy can be seen as the summation of all of the occupied electron energy levels. It is accepted that in most cases the distortion energy to form a bipolaron is roughly equal to that to form two polarons but the ionization energy of bipolaron is lower than two polarons. Thus the stability of bipolaron lies on the competition between electron-phonon interaction and the Coulomb repulsion of two like charges. From the dash lines in Fig. 2, one can find the coulomb repulsion $(U_0, V_0)$ do delocalized the configuration of bipolaron but it is still not strong enough to separate the bounded dication.

Since the screening of the $e$-$e$ interaction is thought to be very important to the bipolaron stability, we study its effect with careful calculations by continuously turning on the screening from zero to its realistic value, that is, we define a screened factor $f_s$ and then let it run over form 0 to 1, $U = (1-f_s)U_0 + f_s U_c$, $V = (1-f_s)V_0 + f_s V_c$. A localized factor is introduced to estimate the localization of the bipolaron state as, $f_l = \sum_n |\Psi_{n,N}|^4$, where $\Psi_{n,N}$ is the wavefunction of bipolaron. For a completed extended state, $f_l \to 0 (N \to \infty)$ while for a completed localized state, $f_l = 1$. In general, we have $0 \le f_l \le 1$. So the value of $f_l$ can indicate the localized degree of electronic states. The dependence of localized factor of the bipolaron state on the screened factor is shown in the insert of Fig. 2b. If the $e$-$e$ interaction is totally unscreened($f_s = 0$), the extending of the bipolaron is about 6 to 8 base pairs. With the increase of the screening, the spatial extension of the bipolaron becomes narrower and the localization is strengthen. The localized factor is found to approximately be saturated at a strong localized state($f_l \sim 0.5$) when the screening increases near that in the water solution($f_s \sim 1$).

For further clarification, we consider the creation energy difference between the bipolaron and two polarons. The creation energy of a defect state, from its uniform counterpart, is defined as $E^{\mathrm{cr}} = E_T^{\mathrm{d}} - E_T^{\mathrm{u}} - \mu(N_e^{\mathrm{d}} - N_e^{\mathrm{u}})$, where $\mu = (e_v + e_c)/2$, with $e_v$ ($e_c$) being the single-particle energy level at the top (bottom) of the valence (conduction) band, while $N_e^{\mathrm{d}}$ ($N_e^{\mathrm{u}}$) denotes the number of electrons in the defect (uniform) state [28]. Thus, the creation energy difference between the bipolaron and two polarons is

$$\Delta E^{\mathrm{cr}} = E_{\mathrm{bp}}^{\mathrm{cr}} - 2E_{\mathrm{p}}^{\mathrm{cr}} = E_T^{\mathrm{bp}} + E_T^{\mathrm{u}} - 2E_T^{\mathrm{p}}. \qquad (9)$$

Shown in Fig. 3 is the $\Delta E^{\mathrm{cr}}$ as a function of the length of the model DNA molecule at screened factor $f_s = 0.8$, a moderate value chosen to not only account for the solvent polarization effects but also hope the results to be reported qualitatively true even under the dry DNA condition. The bipolaron is more stable than its two-polaron counterpart by about 0.5 eV, which is in the same order of that in conducting polymers [17]. The figure's insert is the dependence of $\Delta E^{\mathrm{cr}}$ on the screened factor $f_s$ at $N = 28$, showing it again that the bipolaron is generally more favored by solvent polarization. When $f_s$ increases from 0 to 0.8, $-\Delta E^{\mathrm{cr}}$ is enhanced by about 15%. These results have thus clearly demonstrated that in the DNA molecule a self-trapping bipolaron is more stable than two polarons. They have also emphasized the important role of the screening of the $e$-$e$ interaction on the stability of bipolaron in DNA.

Extensive studies of the dynamics of excitations under the influence of an electric field have been done in conducting polymers. Su and Schrieffer [29] and Phillpot *et al.* [30] considered that polarons can keep their shape while moving along a chain. However, some author showed that it may not be true under high electric field. In that case, the charge moves faster and not allows the distortion occur, thus the polaron may become delocalized [25].

To study the transport properties of bipolaron in DNA, we use the same dynamical technique as in Refs. 26 and 27 to solve the time-dependent Schrödinger equation ($f_s = 0.8$) coupled with the Newtonian equation of motion for the base displacements [25]. Unlike in conducting polymers, bipolaron needs high kinetic energy for its transport in a DNA sequence. We choose the driving electric field of $\mathcal{E} = 20\,\mathcal{E}_0$, which produces a voltage drop of about ~10 V between the two ends of the DNA stack. The evolution of $Q_n^h(t)$ is depicted in Fig. 4. Our results are consistent with the suggestion that the excess charges move long distances through DNA by a series of hopping process [31]. At $t = 0$, a bipolaron is localized at $(G/C)_{14}(G/C)_{15}$, and then the electric field $\mathcal{E}$ smoothly turn on from 0 to $20\,\mathcal{E}_0$ in 0.5 ps. After a relaxation time about 7 ps to overcome the trapping energy, the dication moves along the direction of $E$ to the right end through series trapping and hopping processes. It is found that the wave function of the dication can keep its localization characteristic throughout the hopping process, but the lattice distortion partially loses bipolaron configuration instantaneously because of its delay to the motion of charges. Between every two adjacent hopping processes, we find the dication stays at some base pairs for a time interval before it begins next hop. These base pairs corresponds to the trapping centers for dication which has large negative hole trapping energy [32]. It is confirmed that single or multiple G units can act as the trapping centers in the dication hopping process. The dication stays at $(C/G)_{15}(G/C)_{16}$ for about 5 ps; at $(C/G)_{20}$ for about 7 ps before it makes a further hop to next center(s) or right end.

It has been reported that in a bond-disordered polymer chain the strong electrical field increases the nonadiabatic effects to make the gap states mixed each other[26]. Here we investigate the nonadiabatic effects on the bipolaron state in disordered DNA oligomer by checking the evolution of the occupation number associated with the in-stantaneous bipolaron energy level and plot the result in the insert of Fig. 4. Similarly, the strong electrical field has changed the occupation number of the bipolaron energy level. The nonadiabatic effects are responsible for the bipolaron partially losing its configuration during the hopping. In particular between 13.5 ps to 20 ps, the occupation number of bipolaron level is found very close to 1, which implies a strong mixed state of the initially unoccupied state and the initially double-occupied HOMO state. In this case, the dication is driven by the strong electronic field to more like two polarons than one bipolaron. After 25 ps, the occupation number returns a small value indicating the reformation of a bipolaron near the right boundary of the DNA oligomer, as shown in Fig. 4.

By a rough estimation, the dication can transport 50 Å in 25 ps driven by an electric field $\mathcal{E} = 1.06 \times 10^6$ V/cm. The transport speed can reach about 4∼5 Å /ps, if the long trapping time is not considered, Such a fast dication hopping in DNA may be comparative to the long-range charge transport in experiments, as Wan *et al.* reported that the hole(s) can transport a 17 Å distance in 5 ps by thermal activation [33].

Recently, Kato *et al.* have observed the excited electrons located in the $\pi^*$ states by resonant photoemission spectroscopy (RPES) near the Fermi level [34]. The optical signals are considered to be of polarons or bipolarons in DNA. The detailed experimental study on the bipolarons in DNA should investigate the multi-step redox states with optical and/or electron spin resonance (ESR) spectroscopy by analogy with the experiments in conducting polymers [35]. Important evidence for the generation of bipolarons can also be provided by the loss of the ESR signal at higher doping levels.

In summary we theoretically showed that a bipolaron is more stable than two polarons in DNA in both dry and solution conditions. The dication can move quite a long distance through DNA by a series of hopping process from one trapping center to another, partially losing its configuration instantaneously induced by the nonadiabatic effects. The above conclusion has been testified by the calculations on DNA stacks with different random base pairs sequence and different total sites $N$.

## Acknowledgments

Support from the Research Grants Council of the Hong Kong Government (Grant No. 604804) and the National Natural Science Foundation of China (Grant No. 10474056) are gratefully acknowledged. The authors are grateful to Dr. H. Y. Zhang and Dr. J. Berakdar for helpful discussions

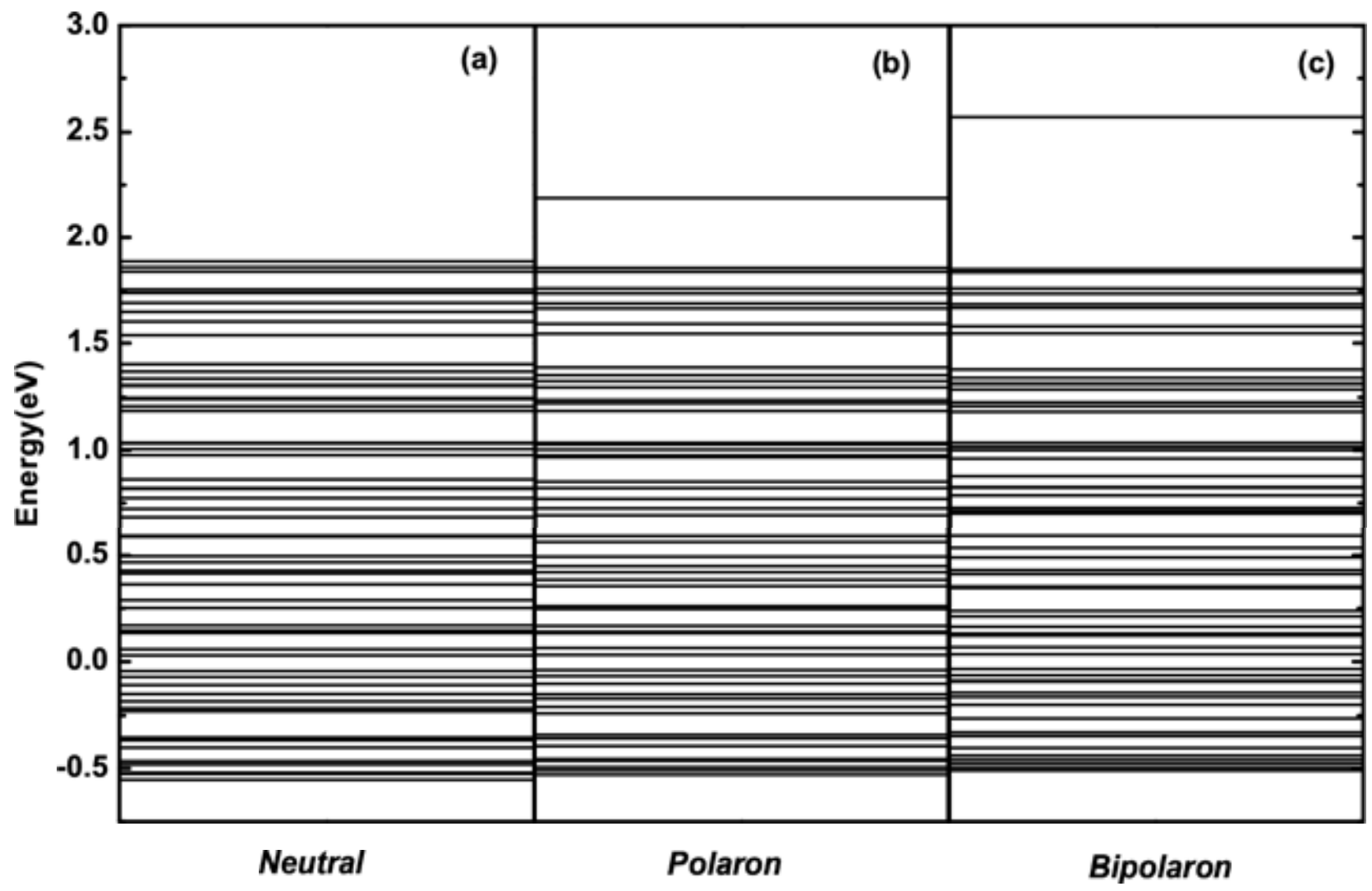

FIG. 1: Energy levels for the random DNA stack with no external field nor $e$-$e$ interaction concerned. (a) neutral state, (b) polaron state induced by one hole doping, and (c) bipolaron state induced by two holes doping. The zero of energy is the on-site energy of the T base pair. The $\pi - \pi*$ gap of DNA is $\sim$ 3.75 eV[3] and the polaron/bipolaron gap level is 0.35/0.72 eV above the highest filled level.

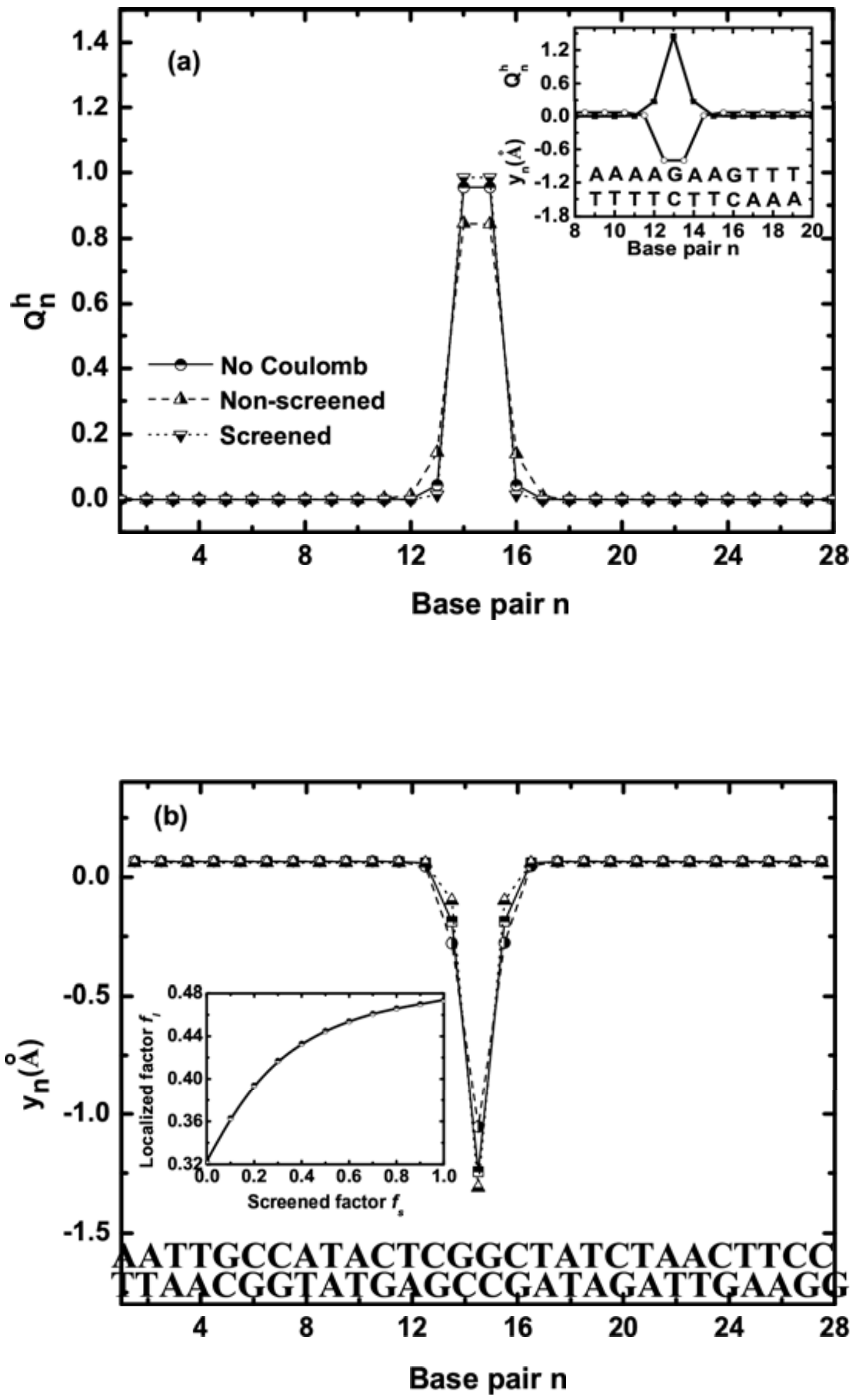

FIG. 2: Dication population (a) and the lattice distortion (b) for the random DNA stack under three conditions, without $e$-$e$ interaction (solid lines), with non-screened (dash lines) and with screened $e$-$e$ interaction (dot lines). The insert in (a) shows the bipolaron self-trapping at a G/C-singlet site in a sequence contains no G/C-multiplet. The insert in (b) shows the favorable of screening to the formation of bipolaron

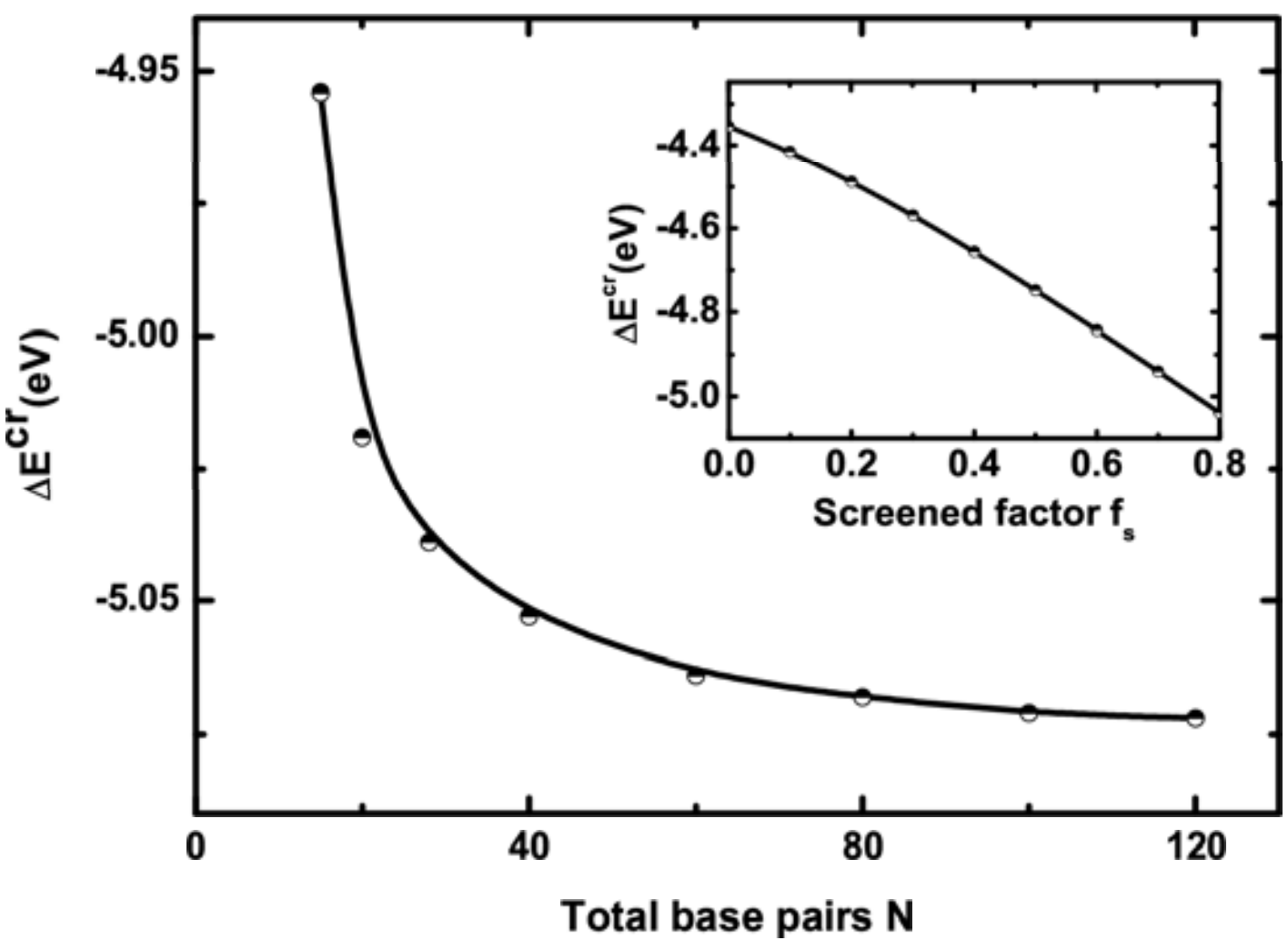

FIG. 3: $\Delta E^{\mathrm{cr}}$ of bipolaron and two polarons as a function of the total site number $N$ at $f_s = 0.8$. The insert shows the dependence of $\Delta E^{\mathrm{cr}}$ on $f_s$ at $N = 28$.

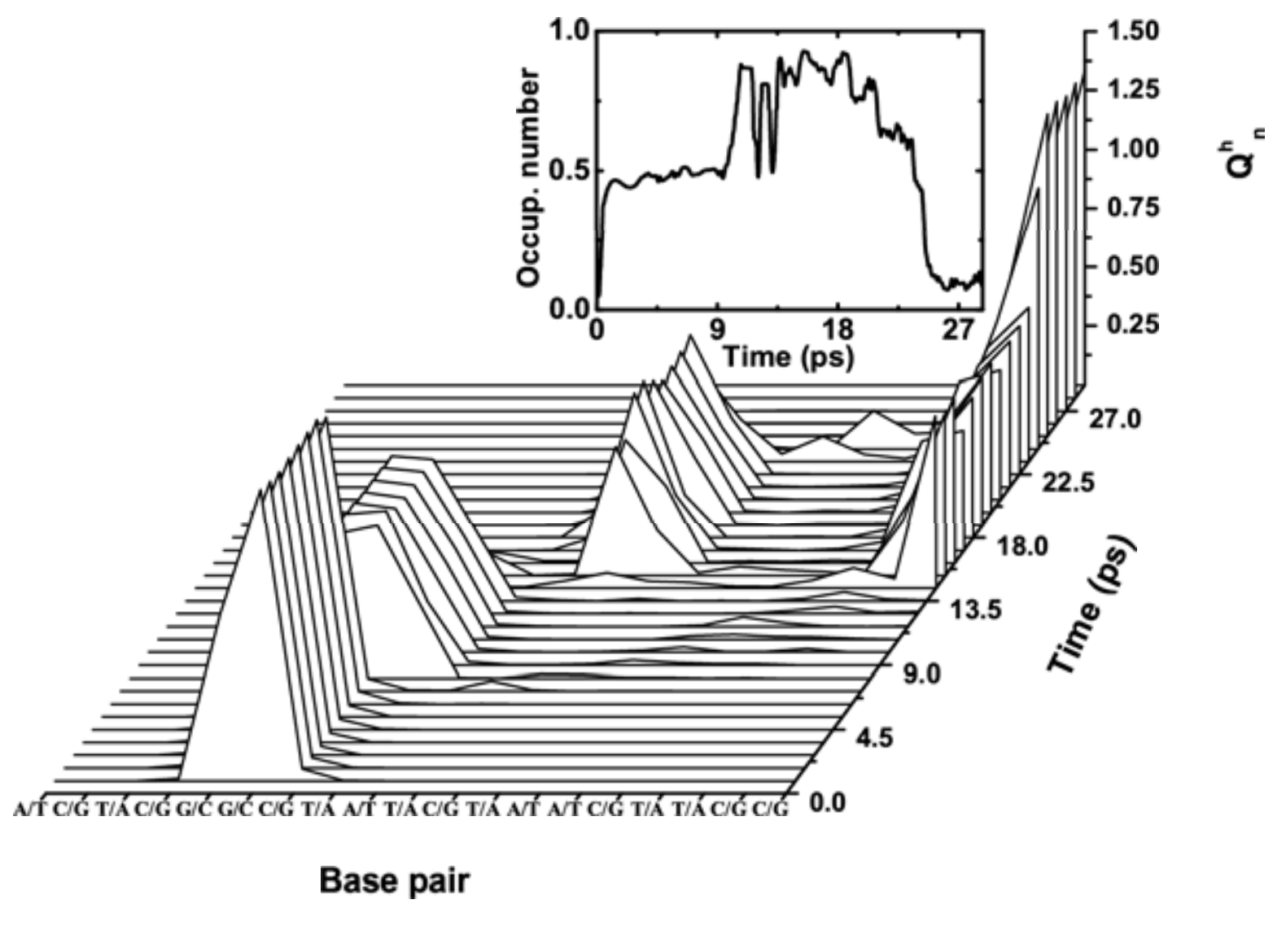

FIG. 4: The excess charges distribution as a function of time and site under electric field $\mathcal{E} = 20\mathcal{E}_0$ at $f_s = 0.8$. The insert shows the evolution of the occupation number of the bipolaron energy level. The occupation number of bipolaron level between 13.5$\sim$20 ps implies a strong mixed gap state due to the nonadiabatic effects under high electrical field.